**Comparative study of semilocal density functionals on solids and surfaces**

Yuxiang Mo[1], Guocai Tian[1,2], and Jianmin Tao[1,*]

[1]Department of Physics, Temple University, Philadelphia, Pennsylvania 19122, USA
[2]State Key Laboratory of Complex Nonferrous Metal Resources Clean Utilization, Kunming University of Science and Technology, Kunming 650093, China

**Abstract**

Recently, Tao and Mo (TM) proposed an accurate all-purpose nonempirical meta-generalized gradient approximation (meta-GGA). The exchange part was derived from the density matrix approximation, while the correlation part is based on a modification of TPSS correlation in the low-density or strong-interaction limit. To further understand this density functional, we combine the TM exchange part with the original TPSS correlation and make a comprehensive assessment of this combination, which we call TMTPSS functional, on solids and solid surfaces. Our test includes 22 lattice constants and bulk moduli, 30 band gaps of semiconductors, 7 cohesive energies, and surface exchange-correlation energies for $r_s$ ranging from 2 to 3 bohr. Our calculations show that TMTPSS functional is quite competitive to the TM meta-GGA functional, improving upon the nonempirical functionals LSDA, PBE GGA, and TPSS meta-GGA for most properties considered. In particular, it significantly improves the surface exchange-correlation energy calculation, with a mean absolute error of only 1 erg/cm$^2$.

1. **INTRODUCTION**

Because of the good balance between computational cost and achievable accuracy, Kohn-Sham (KS) density functional theory [1] (DFT) is the most popular method in electronic structure calculations of molecules and solids. Since the exact kinetic energy and classical Coulomb interaction energy can be expressed explicitly in terms of the KS single-particle orbitals, the key to the improvement of the theory for achieving higher accuracy and versatility is to approximate the interaction between an electron and the associated exchange-correlation hole around the electron.

According to the type of local ingredients, density functionals can be divided into two broad categories, regardless of how they are developed. One is called semilocal DFT [2–14] and the other is nonlocal DFT. [15,16] The former are developed using the semilocal information, such as spin electronic densities, density derivatives, and the orbital kinetic energy density, while the latter make use of not only the semilocal information, but also the nonlocal information, including the exact exchange energy density, [17–19] and in some cases, even the unoccupied KS orbitals. Clearly, the latter contain much more information and are often more accurate than semilocal approximations, in particular for some properties related to electronic nonlocality.[5] However, they are more complicated and difficult to develop and implement, and thus are less popular than semilocal DFT.

Semilocal density functionals are usually designed to satisfy certain known exact constraints (e.g., those listed in Ref. 21). All or part of the parameters introduced in a particular density functional can be fitted to the properties either experimentally measured or calculated with high-level electronic structure theory. Density functionals developed following this approach are called semiempirical or empirical. They usually have amazingly high accuracy for a target group of systems and properties. But the accuracy can plunge precipitously when such a functional is applied to study physically different systems such as atoms, molecules, solids, and solid surfaces. Examples of semiempirical functionals include BLYP,[3,15,22] B3LYP,[23,24] and M06.[25] Another method of developing exchange-correlation functionals is fitting to real many-electron systems. In comparison with functionals empirically fitted to a set of properties or systems, nonempirical functionals have more consistent performance for diverse systems and properties, though they may not provide the same level of accuracy for certain family of systems. The universality of such nonempirical functionals is particularly useful in situations where the studied systems involve a combination of molecules and solids/surfaces (e.g., chemical reactions on surfaces).

Recently, Tao and Mo (TM) have proposed a meta-GGA functional. All the parameters introduced in this functional are determined by the exact constraints, rather than empirical fitting. Unlike many semilocal functionals, for which the underlying exchange-correlation holes are unknown or constructed with reverse-engineering approach, TM functional is directly obtained from a semilocal exchange-correlation hole. In the present work, we therefore assess the performance of the TM exchange functional when combined with the TPSS correlation functional on lattice constants, bulk moduli, semiconductor bandgaps, cohesive energies, and surface exchange and correlation energies of jellium. For convenience, we call this combination TMTPSS functional. Our calculations show that TMTPSS functional is competitive with TM functional on many properties. In particular, it substantially improves upon all semilocal DFT for surface exchange-correlation energy.

## 2. COMPUTATIONAL METHOD

We assess the TMTPSS functional on 22 bulk crystalline solids including main-group metals Li, K, Al, semiconductors diamond, Si, $\beta$-SiC, Ge, BP, AlP, AlAs, GaN, GaP, GaAs, ionic crystals NaCl, NaF, LiCl, LiF, MgO, MgS, and transition metals Cu, Pd, Ag. A locally modified version [26] of the Gaussian program [27] with periodic boundary conditions (PBC) [28] was used to evaluate the TMTPSS functional on these solids. We adopt the Gaussian-type basis sets which are described in Ref. 29. We used sufficiently dense $k$-point meshes for reliable evaluation of energy: $22\times22\times22$-$20\times40\times40$ for main group metals, $10\times10\times10$-$12\times12\times12$ for semiconductors, $10\times10\times10$-$14\times14\times14$ for ionic crystals, and $8\times16\times16$-$10\times18\times18$ for transition metals. An increase in $k$-point density from these settings does not produce nontrivially different results. We calculate total energies of no less than 10 static lattices with cell volumes ranging from -5% to +5% of the equilibrium cell volume. Such a setting adequately maps out the energy versus volume around the minimum energy and enables the calculation of equilibrium lattice constants and bulk moduli. The distribution of the volumetric data points within the range, either uniform or denser near the energy minimum, was found to have negligible influence on the outcome of lattice constants and bulk moduli. The obtained energy dependence on unit cell volume was then fitted to the stabilized jellium equation of state (SJEOS) [30,31] to generate the equilibrium lattice constants and bulk moduli. Additionally, we evaluate the TMTPSS functional on the bandgaps of 30 semiconductors: C, Si, Ge, SiC, BP, BAs, AlP, AlAs, AlSb, GaN, $\beta$-GaN, GaP, GaAs, GaSb, InN, InP, InAs, InSb, ZnS, ZnSe, ZnTe, CdS, CdSe, CdTe, MgS, MgSe, MgTe, BaS, BaSe, and BaTe. In these calculations we used the basis sets and effective core potentials from the supporting information of Ref. 32.

## 3. RESULTS AND DISCUSSION

### 3.1 Lattice Constants

Accurate prediction of equilibrium lattice constants of a solid is of paramount importance because it is a starting point for exploration of all other properties of the solid. Listed in Table I are the equilibrium lattice constants of 22 solids calculated with the TMTPSS functional along with results of other functionals from literature. Lattice constants of Ge, BP, AlP, AlAs, GaN, GaP, and MgS for LSDA, PBE, TPSS, PBEsol, and revTPSS are from Ref. 33. The other values of LSDA, PBE, and TPSS are from Ref. 21, while those of PBEsol are from Ref. 12. The results of revTPSS are taken from Ref. 13 except that of potassium which is from Ref. 33. The experimental data of lattice constants are quoted from Ref. 21 except those of BP, AlP, AlAs, GaN, GaP, and MgS which are from Ref. 34. All values of TM are from Ref. 29. The values of TMTPSS are calculated self-consistently using Gaussian 09. The TMTPSS functional has a mean absolute error (MAE) of 0.019 Å which is larger than that of TM (MAE = 0.017 Å) but much smaller than those of the LSDA (MAE = 0.062 Å), PBE (MAE = 0.053 Å), and TPSS (MAE = 0.037 Å).

### 3.2 Bulk Moduli

Listed in Table II are the equilibrium bulk moduli of the 22 solids calculated with the TMTPSS functional in comparison with results of other functionals from literature. The TMTPSS functional has an MAE of 6.9 GPa, which is slightly smaller than that of TM (MAE = 7.0 GPa), and significantly smaller than those of TPSS (MAE = 8.8 GPa), PBE (MAE = 7.8 GPa), and LSDA (MAE = 12.0 GPa). The TMTPSS functional has higher accuracy than TM for two of the main group metals Li and Al, while predicting the same bulk modulus for K. For semiconductors, the TMTPSS functional yields less accurate bulk moduli than those of TM for diamond, Si, and SiC, but it has a higher accuracy on the bulk modulus of GaAs. The TMTPSS and TM functionals produce the same bulk moduli to the first decimal place for the rest six semiconductors. In terms of ionic solids, the TMTPSS functional is more accurate than TM on LiCl, LiF, and MgO, while less accurate on the other three. When it comes to the three transition metals, the TMTPSS functional predicts better bulk moduli than TM for Pd and Ag, but a less accurate one for Cu.

**Table I**: Equilibrium lattice constants (Å) of 22 solids at 0 K calculated from the fitting of SJEOS. The smallest and largest MAEs are in bold blue and red, respectively.

| Solids | Expt. | LSDA | PBE | TPSS | TM | TMTPSS |
|---|---|---|---|---|---|---|
| Li | 3.477 | 3.383 | 3.453 | 3.475 | 3.445 | 3.445 |
| K | 5.225 | 5.093 | 5.308 | 5.362 | 5.265 | 5.270 |
| Al | 4.032 | 4.008 | 4.063 | 4.035 | 4.024 | 4.026 |
| C | 3.567 | 3.544 | 3.583 | 3.583 | 3.564 | 3.570 |
| Si | 5.430 | 5.426 | 5.490 | 5.477 | 5.443 | 5.456 |
| SiC | 4.358 | 4.351 | 4.401 | 4.392 | 4.374 | 4.381 |
| Ge | 5.652 | 5.624 | 5.764 | 5.723 | 5.671 | 5.671 |
| BP | 4.538 | 4.491 | 4.548 | 4.544 | 4.534 | 4.534 |
| AlP | 5.460 | 5.433 | 5.504 | 5.492 | 5.487 | 5.487 |
| AlAs | 5.658 | 5.631 | 5.728 | 5.702 | 5.691 | 5.691 |
| GaN | 4.531 | 4.457 | 4.549 | 4.532 | 4.492 | 4.492 |
| GaP | 5.448 | 5.392 | 5.506 | 5.488 | 5.437 | 5.437 |
| GaAs | 5.648 | 5.592 | 5.726 | 5.702 | 5.641 | 5.660 |
| NaCl | 5.595 | 5.471 | 5.698 | 5.696 | 5.618 | 5.637 |
| NaF | 4.609 | 4.505 | 4.700 | 4.706 | 4.626 | 4.633 |
| LiCl | 5.106 | 4.968 | 5.148 | 5.113 | 5.089 | 5.112 |
| LiF | 4.010 | 3.904 | 4.062 | 4.026 | 3.995 | 4.003 |
| MgO | 4.207 | 4.156 | 4.242 | 4.224 | 4.209 | 4.218 |
| MgS | 5.202 | 5.127 | 5.228 | 5.228 | 5.198 | 5.197 |
| Cu | 3.603 | 3.530 | 3.636 | 3.593 | 3.587 | 3.578 |
| Pd | 3.881 | 3.851 | 3.950 | 3.917 | 3.900 | 3.909 |
| Ag | 4.069 | 3.997 | 4.130 | 4.076 | 4.052 | 4.064 |
| ME | | -0.062 | 0.051 | 0.035 | 0.002 | 0.008 |
| MAE | | **0.062** | 0.053 | 0.037 | **0.017** | 0.019 |

**Table II**: Equilibrium 0K bulk moduli (GPa) of the 22 solids calculated from the fitting of SJEOS. The LSDA, PBE, and TPSS values are from Ref. 21. For BP, AlP, AlAs, GaN, GaP, and MgS, the LSDA and PBE values are taken from Ref. 35. All values of TM are from Ref. 29. The experimental values used for error calculation are from these references: Li,[36] K,[37] Al,[38] C,[39] Si,[40] SiC,[41] Ge,[40] BP,[42] AlP,[43] AlAs,[43] GaN,[44] GaP,[43] GaAs,[40] NaCl,[45] NaF,[45] LiCl,[45] LiF,[46] MgO,[47] MgS,[48] Cu,[49] Pd,[50] and Ag.[51] The smallest and largest MAEs are in bold blue and red, respectively.

| Solids | LSDA | PBE | TPSS | TM | TMTPSS | Expt. |
|---|---|---|---|---|---|---|
| Li | 14.7 | 13.7 | 13.2 | 13.7 | 13.5 | 13 |
| K | 4.6 | 3.8 | 3.6 | 4.0 | 4.0 | 3.7 |
| Al | 82.5 | 76.8 | 85.2 | 88.6 | 86.9 | 79.4 |
| C | 458 | 426 | 421 | 442.4 | 435.5 | 443 |
| Si | 95.6 | 89 | 91.9 | 97.1 | 95.1 | 99.2 |
| SiC | 225 | 209 | 213 | 220.0 | 217.5 | 225 |
| Ge | 75.9 | 63.0 | 66.4 | 72.5 | 72.5 | 75.8 |
| BP | 176 | 162 |  | 171.5 | 171.5 | 173 |
| AlP | 89.9 | 82.6 |  | 89.3 | 89.3 | 86 |
| AlAs | 75.5 | 67.0 |  | 75.2 | 75.2 | 82 |
| GaN | 204 | 173 |  | 207.1 | 207.1 | 190 |
| GaP | 90.6 | 77.0 |  | 89.2 | 89.2 | 88 |
| GaAs | 81.3 | 68.1 | 70.1 | 78.6 | 75.9 | 75.6 |
| NaCl | 32.5 | 23.9 | 23 | 26.9 | 26.2 | 26.6 |
| NaF | 63.3 | 47.7 | 44 | 52.5 | 52.7 | 51.4 |
| LiCl | 42 | 32.9 | 34.3 | 36.2 | 35.3 | 35.4 |
| LiF | 87.5 | 65.9 | 67.2 | 74.4 | 74.3 | 69.8 |
| MgO | 183 | 162 | 169 | 174.5 | 171.4 | 165 |
| MgS | 84.0 | 74.4 |  | 79.8 | 80.3 | 78.9 |
| Cu | 192 | 153 | 173 | 180.2 | 181.3 | 142 |
| Pd | 240 | 180 | 203 | 210.7 | 206.0 | 195 |
| Ag | 153 | 107 | 129 | 138.4 | 134.6 | 109 |
| ME | 11.1 | -6.8 | -0.1 | 5.3 | 4.0 |  |
| MAE | **12.0** | 7.8 | 8.8 | 7.0 | **6.9** |  |
| MARE | 0.1 | 0.1 | 0.1 | 0.1 | 0.1 |  |

### 3.3 Semiconductor Band Gaps

The bandgap is a deciding quantity for electrical, optical, photoelectric, and photocatalytic characteristics of semiconductors. Listed in Table III are the results on bandgaps for 30 semiconductors. The experimental values are from Ref. 32. The LSDA, PBE, PBEsol, TPSS, and revTPSS values are from Ref. 52. Similar to LSDA, GGAs, and meta-GGAs listed in Table III, the TMTPSS functional underestimates the bandgap for every one of the 30 semiconductors, as indicated by the equal magnitude of ME and MAE. The MAE (0.89 eV) of the TMTPSS functional is larger than that of TPSS (MAE=0.79 eV) but smaller than those of LSDA (MAE=1.09 eV), PBE (MAE=0.93 eV), PBEsol (MAE=1.09 eV), and revTPSS (MAE=0.95 eV). Since the predictions of bandgap and lattice constant often appear as a tradeoff, the accuracy of the TMTPSS functional on bandgaps is satisfactory, considering the remarkably accurate lattice constant this functional can yield. The TMTPSS functional predicts zero bandgap for one semiconductor –InN. It is the only functional which predicts zero bandgap for only one semiconductor among the 30, in contrast to LSDA (5 zero bandgaps), PBE (3 zero bandgaps), PBEsol (4 zero bandgaps), TPSS (2 zero bandgaps), and revTPSS (2 zero bandgaps). For these difficult cases of bandgap predictions, the TMTPSS functional also has the best estimate of the numerical values of the gaps, significantly better than the other functionals listed.

**Table III**: Bandgaps (in eV) of 30 semiconductors. The smallest and largest MAEs are in bold blue and red, respectively.

| Solid | Expt. | LSDA | PBE | PBEsol | TPSS | revTPSS | TMTPSS |
|---|---|---|---|---|---|---|---|
| C | 5.48 | 4.22 | 4.24 | 4.03 | 4.29 | 4.05 | 4.15 |
| Si | 1.17 | 0.62 | 0.72 | 0.53 | 0.80 | 0.63 | 0.65 |
| Ge | 0.74 | 0.00 | 0.13 | 0.00 | 0.32 | 0.14 | 0.28 |
| SiC | 2.42 | 1.42 | 1.46 | 1.27 | 1.42 | 1.23 | 1.31 |
| BP | 2.40 | 1.36 | 1.40 | 1.24 | 1.45 | 1.28 | 1.34 |
| BAs | 1.46 | 1.19 | 1.25 | 1.10 | 1.27 | 1.13 | 1.17 |
| AlP | 2.51 | 1.64 | 1.78 | 1.56 | 1.86 | 1.72 | 1.76 |
| AlAs | 2.23 | 1.43 | 1.55 | 1.37 | 1.66 | 1.57 | 1.60 |
| AlSb | 1.68 | 1.34 | 1.44 | 1.22 | 1.58 | 1.40 | 1.37 |
| GaN | 3.50 | 2.18 | 2.22 | 1.85 | 2.15 | 1.71 | 1.58 |
| $\beta$-GaN | 3.30 | 1.84 | 1.86 | 1.70 | 1.79 | 1.53 | 1.75 |
| GaP | 2.35 | 1.63 | 1.80 | 1.62 | 1.89 | 1.77 | 1.74 |
| GaAs | 1.52 | 0.04 | 0.36 | 0.42 | 0.60 | 0.73 | 0.83 |
| GaSb | 0.73 | 0.00 | 0.19 | 0.06 | 0.39 | 0.31 | 0.43 |
| InN | 0.69 | 0.00 | 0.00 | 0.00 | 0.00 | 0.01 | 0.00 |
| InP | 1.42 | 0.74 | 0.99 | 0.83 | 1.19 | 1.00 | 1.07 |
| InAs | 0.41 | 0.00 | 0.00 | 0.00 | 0.08 | 0.00 | 0.06 |
| InSb | 0.23 | 0.00 | 0.00 | 0.00 | 0.00 | 0.00 | 0.03 |
| ZnS | 3.66 | 2.02 | 2.30 | 2.22 | 2.53 | 2.42 | 2.42 |
| ZnSe | 2.70 | 1.05 | 1.37 | 1.26 | 1.62 | 1.58 | 1.62 |
| ZnTe | 2.38 | 1.11 | 1.39 | 1.29 | 1.65 | 1.60 | 1.67 |
| CdS | 2.55 | 0.97 | 1.26 | 1.08 | 1.47 | 1.31 | 1.29 |
| CdSe | 1.90 | 0.31 | 0.63 | 0.45 | 0.85 | 0.77 | 0.81 |
| CdTe | 1.92 | 0.54 | 0.81 | 0.67 | 1.05 | 0.98 | 1.05 |
| MgS | 5.40 | 3.37 | 3.65 | 3.34 | 3.91 | 2.68 | 3.71 |
| MgSe | 2.47 | 1.74 | 1.90 | 1.70 | 2.21 | 2.03 | 1.98 |
| MgTe | 3.60 | 2.41 | 2.65 | 2.58 | 3.07 | 3.08 | 3.10 |
| BaS | 3.88 | 2.13 | 2.40 | 2.15 | 2.56 | 2.48 | 2.46 |
| BaSe | 3.58 | 1.84 | 2.05 | 1.83 | 2.18 | 2.17 | 2.18 |
| BaTe | 3.08 | 1.48 | 1.66 | 1.38 | 1.77 | 1.69 | 1.71 |
| ME | | -1.09 | -0.93 | -1.09 | -0.79 | -0.95 | -0.87 |
| MAE | | **1.09** | 0.93 | **1.09** | **0.79** | 0.95 | 0.87 |

### 3.4 Cohesive Energies

Cohesive energy is the energy needed for the destruction of interatomic bonds and completely freeing atoms from each other. In practice, we first calculate the electronic energy per atom within the solid. This energy is then added with the phonon zero-point energy. The zero-point energy (a result of the zero-point motion) per atom is approximated [30] with Debye temperatures from experiments: C 2230K,[53] Si 645K,[53] SiC 1232K,[54] NaCl 321K,[53] NaF 492K,[53] LiCl 422K,[53] and LiF 732K.[53] The cohesive energy is the difference between the ZPE-corrected energy per atom and the spin-polarized ground-state energy of the atom in free space. Molecular basis sets of elements Li and Na include diffuse functions, which are excluded in the calculations of solids to avoid significant computational slow-down from the Coulomb term of the total energy. But for the evaluation of ground-state energies of atoms in free space, the diffuse functions are necessary for the representation of the tail regions of atoms. In cases of Li and Na atoms in free space, we use the full molecular basis set 6-311G*. Here, applying two different basis sets to the same atom in two different environments (in solid vs. in free space) can yield reliable cohesive energies for two reasons. First, diffuse functions only make a trivial contribution to the total energy of the lattice of a solid. Therefore, removing them does not jeopardize the evaluation of the total energy. Furthermore, in those ionic solids, Li and Na cations are electron-shy which makes density tail regions less important. Listed in Table IV are the cohesive energies of the aforementioned 7 solids. The TMTPSS functional is more accurate in terms of cohesive energies than LSDA and TPSS. But the TMTPSS functional has a mean absolute error (MAE) of 0.13 eV/atom, larger than those of the meta-GGA TM (MAE=0.08 eV/atom) and the GGA PBE (MAE=0.12 eV/atom).

**Table IV**: Cohesive energies (eV/atom) of 7 solids. The results of LSDA, PBE, and TPSS are from Ref. 21. Those for TM are from Ref. 29. The values of TMTPSS are corrected for zero-point vibrations. The smallest and largest MAEs are in bold blue and red, respectively.

| Solid | LSDA | PBE | TPSS | TM | TMTPSS | Expt. |
|---|---|---|---|---|---|---|
| C | 8.83 | 7.62 | 7.12 | 7.48 | 7.21 | 7.37 |
| Si | 5.26 | 4.50 | 4.36 | 4.61 | 4.42 | 4.62 |
| SiC | 7.25 | 6.25 | 6.02 | 6.29 | 6.08 | 6.37 |
| NaCl | 3.58 | 3.16 | 3.18 | 3.19 | 3.21 | 3.31 |
| NaF | 4.50 | 3.96 | 3.87 | 3.88 | 3.91 | 3.93 |
| LiCl | 3.88 | 3.41 | 3.41 | 3.42 | 3.44 | 3.55 |
| LiF | 5.02 | 4.42 | 4.32 | 4.34 | 4.36 | 4.40 |
| ME | 0.68 | -0.03 | -0.18 | -0.05 | -0.13 | |
| MAE | **0.68** | 0.12 | 0.18 | **0.08** | 0.13 | |
| MARE | 13.4 | 2.5 | 3.7 | 1.9 | 2.7 | |

### 3.4 Surface Exchange-Correlation Energy

Jellium is a quantum mechanical model in which both positive charges and electron density are uniformly distributed. It is a realistic model for simple metals. Researching the accuracy of a density functional on jellium surface energies can provide useful insight on how the functional would perform on surface calculation of real systems. The distribution of electronic density in the jellium surface model poses a requirement of good behavior in both the slowly varying and rapidly varying regimes to a density functional. We use RPA (random-phase approximation) calculation from $r_s = 2$ bohrs to $r_s = 3$ bohrs as our reference for error analysis, due to the uncertainty in quantum Monte Carlo calculation of jellium surface exchange-correlation energies. Listed in Table V are the jellium surface exchange and correlation energies. The surface exchange-correlation energy of the TMTPSS is surprisingly accurate in reference to the exact values. It has a tiny MAE of 1 erg/cm$^2$ which is in huge contrast to LSDA (MAE=77 erg/cm$^2$), PBE (MAE=133 erg/cm$^2$), and TPSS (MAE=60 erg/cm$^2$), and even TM (MAE=35 erg/cm$^2$).

**Table V**: Jellium surface exchange energies $\sigma_x$ and surface exchange-correlation energies $\sigma_{xc}$ (in erg/cm$^2$). The RPA values are taken from Ref. 55. The LSDA, PBE, and TPSS values are from Ref. 21. Those for TM are from Ref. 29. The smallest and largest MAEs are in bold blue and red, respectively.

| | Exchange | | | | | Exchange-correlation | | | | | |
|---|---|---|---|---|---|---|---|---|---|---|---|
| $r_s$ (bohr) | LSDA | PBE | TPSS | TM | RPA | LSDA | PBE | TPSS | TM | TMTPSS | RPA |
| 2.00 | 3037 | 2438 | 2553 | 2641 | 2624 | 3354 | 3265 | 3380 | 3515 | 3465 | 3467 |
| 2.07 | 2674 | 2127 | 2231 | 2312 | 2296 | 2961 | 2881 | 2985 | 3109 | 3063 | 3064 |
| 2.30 | 1809 | 1395 | 1469 | 1531 | 1521 | 2019 | 1962 | 2035 | 2132 | 2095 | 2098 |
| 2.66 | 1051 | 770 | 817 | 860 | 854 | 1188 | 1152 | 1198 | 1267 | 1239 | 1240 |
| 3.00 | 669 | 468 | 497 | 528 | 526 | 764 | 743 | 772 | 823 | 801 | 801 |
| ME | 284 | -125 | -51 | 10 | | -77 | -133 | -60 | 35 | -1 | |
| MAE | **284** | 125 | 51 | **10** | | 77 | **133** | 60 | 35 | **1** | |

## 4. CONCLUSIONS

In summary, we made an assessment of the accuracy of the TMTPSS functional which is a combination of the TM exchange part and the TPSS correlation part on the lattice constants and bulk moduli of 22 bulk solids, band gaps of 30 semiconductors, cohesive energies of 7 solids, and jellium surface exchange-correlation energies. According to our calculations, the TMTPSS functional is highly accurate in determining jellium surface energies, with a minimal difference from those by random-phase approximation which is computationally much more expensive. This suggests that the TMTPSS functional can be an excellent candidate for accurate and cost-effective DFT studies of metal surfaces and adsorptions. The bulk moduli calculated with TMTPSS are slightly more accurate than those calculated with the TM functional. The TMTPSS functional predicts zero bandgap for only one semiconductor of the 30, which is the fewest among all functionals considered. Therefore, it is a premium choice for distinguishing small-gap semiconductors from conductors. The TMTPSS functional has a slightly larger error than that of TM for lattice constants but is still very accurate in comparison with other functionals under study. It terms of cohesive energies, TMTPSS is less accurate than TM, comparable to PBE, and more accurate than LSDA and TPSS.

**ACKNOWLEDGMENTS**

YM and JT acknowledge support from NSF under Grant no. CHE-1640584. JT also acknowledges support from Temple University. GT was supported by China Scholarship Council and the National Natural Science Foundation of China under Grant No. 51264021. Computational support was provided by HPC of Temple University.

\* Corresponding author. jianmin.tao@temple.edu

**References:**
[1] W. Kohn and L. J. Sham, Phys. Rev. **140**, A1133 (1965).
[2] J. P. Perdew and W. Yue, Phys. Rev. B **33**, 8800 (1986).
[3] A. D. Becke, Phys. Rev. A **38**, 3098 (1988).

4 A. D. Becke and M. R. Roussel, Phys. Rev. A **39**, 3761 (1989).
5 A. D. Becke, J. Chem. Phys. **104**, 1040 (1996).
6 J. P. Perdew, K. Burke, and M. Ernzerhof, Phys. Rev. Lett. **77**, 3865 (1996).
7 T. Van Voorhis and G. E. Scuseria, J. Chem. Phys. **109**, 400 (1998).
8 F. A. Hamprecht, A. J. Cohen, D. J. Tozer, and N. C. Handy, J. Chem. Phys. **109**, 6264 (1998).
9 R. Armiento and A. E. Mattsson, Phys. Rev. B **72**, 085108 (2005).
10 Y. Zhao and D. G. Truhlar, J. Chem. Phys. **125**, 194101 (2006).
11 J. Tao, J. P. Perdew, V. N. Staroverov, and G. E. Scuseria, Phys. Rev. Lett. **91**, 146401 (2003).
12 J. P. Perdew, A. Ruzsinszky, G. I. Csonka, O. A. Vydrov, G. E. Scuseria, L. A. Constantin, X. Zhou, and K. Burke, Phys. Rev. Lett. **100**, 136406 (2008).
13 J. P. Perdew, A. Ruzsinszky, G. I. Csonka, L. A. Constantin, and J. Sun, Phys. Rev. Lett. **103**, 026403 (2009).
14 J. Sun, A. Ruzsinszky, and J. P. Perdew, Phys. Rev. Lett. **115**, 036402 (2015).
15 C. Lee, W. Yang, and R. G. Parr, Phys. Rev. B **37**, 785 (1988).
16 C. Adamo and V. Barone, J. Chem. Phys. **110**, 6158 (1999).
17 J. Jaramillo, G. E. Scuseria, and M. Ernzerhof, J. Chem. Phys. **118**, 1068 (2003).
18 A. V. Arbuznikov and M. Kaupp, J. Chem. Phys. **128**, 214107 (2008).
19 M. A. L. Marques, J. Vidal, M. J. T. Oliveira, L. Reining, and S. Botti, Phys. Rev. B **83**, 035119 (2011).
20 J. P. Perdew, V. N. Staroverov, J. Tao, and G. E. Scuseria, Phys. Rev. A **78**, 052513 (2008).
21 V. N. Staroverov, G. E. Scuseria, J. Tao, and J. P. Perdew, Phys. Rev. B **69**, 075102 (2004).
22 B. Miehlich, A. Savin, H. Stoll, and H. Preuss, Chem. Phys. Lett. **157**, 200 (1989).
23 P. J. Stephens, F. J. Devlin, C. F. Chabalowski, and M. J. Frisch, J. Phys. Chem. **98**, 11623 (1994).
24 R. H. Hertwig and W. Koch, Chem. Phys. Lett. **268**, 345 (1997).
25 Y. Zhao and D. G. Truhlar, Theor. Chem. Acc. **120**, 215 (2008).
26 J. Tao and Y. Mo, Phys. Rev. Lett. **117**, 073001 (2016).
27 M. J. Frisch *et al*, *Gaussian 09, Revision A.02* (Gaussian, Inc., Wallingford CT, 2009).
28 K. N. Kudin and G. E. Scuseria, Phys. Rev. B **61**, 16440 (2000).
29 Y. Mo, R. Car, V. N. Staroverov, G. E. Scuseria, and J. Tao, submitted. http://arxiv.org/abs/1607.05252
30 A. B. Alchagirov, J. P. Perdew, J. C. Boettger, R. C. Albers, and C. Fiolhais, Phys. Rev. B **63**, 224115 (2001).
31 A. B. Alchagirov, J. P. Perdew, J. C. Boettger, R. C. Albers, and C. Fiolhais, Phys. Rev. B **67**, 026103 (2003).
32 J. Heyd, J. E. Peralta, G. E. Scuseria, and R. L. Martin, J. Chem. Phys. **123**, 174101 (2005).
33 P. Hao, Y. Fang, J. Sun, G. I. Csonka, P. H. T. Philipsen, and J. P. Perdew, Phys. Rev. B **85**, 014111 (2012).
34 P. Haas, F. Tran, and P. Blaha, Phys. Rev. B **79**, 085104 (2009).
35 F. Tran, R. Laskowski, P. Blaha, and K. Schwarz, Phys. Rev. B **75**, 115131 (2007).
36 R. A. Felice, J. Trivisonno, and D. E. Schuele, Phys. Rev. B **16**, 5173 (1977).
37 W. R. Marquardt and J. Trivisonno, J. Phys. Chem. Solids **26**, 273 (1965).
38 G. N. Kamm and G. A. Alers, J. Appl. Phys. **35**, 327 (1964).
39 M. Levinshtein, S. Rumyantsev, and M. Shur, *Handbook Series on Semiconductor Parameters* (World Scientific, Singapore, 1996).
40 K.-H. Hellwege, *Landolt–Bornstein, New Series, Group III* (Springer, Berlin, 1966).

[41] W. R. L. Lambrecht, B. Segall, M. Methfessel, and M. van Schilfgaarde, Phys. Rev. B **44**, 3685 (1991).
[42] S. Kalvoda, B. Paulus, P. Fulde, and H. Stoll, Phys. Rev. B **55**, 4027 (1997).
[43] *Semiconductors. Physics of Group IV Elements and III-V Compounds.* (Springer-Verlag GmbH, 1982).
[44] H. M. Tütüncü, S. Bağci, G. P. Srivastava, A. T. Albudak, and G. Uğur, Phys. Rev. B **71**, 195309 (2005).
[45] J. T. Lewis, A. Lehoczky, and C. V. Briscoe, Phys. Rev. **161**, 877 (1967).
[46] C. V. Briscoe and C. F. Squire, Phys. Rev. **106**, 1175 (1957).
[47] K. Marklund and S. A. Mahmoud, Phys. Scr. **3**, 75 (1971).
[48] S. M. Peiris, A. J. Campbell, and D. L. Heinz, J. Phys. Chem. Solids **55**, 413 (1994).
[49] W. C. Overton and J. Gaffney, Phys. Rev. **98**, 969 (1955).
[50] D. K. Hsu and R. G. Leisure, Phys. Rev. B **20**, 1339 (1979).
[51] J. R. Neighbours and G. A. Alers, Phys. Rev. **111**, 707 (1958).
[52] R. Peverati and D. G. Truhlar, J. Chem. Phys. **136**, 134704 (2012).
[53] D. E. Gray, *American Institute of Physics Handbook* (McGraw–Hill, New York, 1972).
[54] A. Zywietz, K. Karch, and F. Bechstedt, Phys. Rev. B **54**, 1791 (1996).
[55] J. M. Pitarke and A. G. Eguiluz, Phys. Rev. B **63**, 045116 (2001).